\documentstyle[12pt]{article}

\begin {document}
\begin{flushright} 
OITS 584\\
\vspace{-.5cm}
September 1995
\end{flushright}
\vspace*{.5cm}

\begin{center} {\large {\bf PHASE TRANSITION IN EVOLUTIONARY GAMES}}
\vskip .5cm
 {\bf  Zhen CAO and Rudolph C. HWA} 
\vskip.3cm
 {Institute of Theoretical Science and Department of Physics\\  University of
Oregon, Eugene, OR 97403}
\end{center}

\vskip .5cm

\begin{abstract}

The evolution of cooperative behaviour is studied in the deterministic version of
the Prisoners' Dilemma on a two-dimensional lattice.  The payoff parameter is set
at the critical region $1.8 < b < 2.0$ , where clusters of cooperators are formed in all
spatial sizes.  Using the factorial moments developed in particle and nuclear
physics for the study of phase transition, the distribution of cooperators is studied
as a function of the bin size covering varying numbers of lattice cells.  From the
scaling behaviour of the moments a scaling exponent is determined and is found to
lie in the range where phase transitions are known to take place in physical
systems.  It is therefore inferred that when the payoff parameter is increased
through the critical region the biological system of cooperators undergoes a phase
transition to defectors.  The universality of the critical behaviour is thus extended
to include also this particular model of evolution dynamics. 
\end{abstract}

Phase transition in
evolutionary biology is an important concept.  It is not new.  For example, the
emergence of collectively autocatalytic closure in sufficiently complex sets of
catalytic polymers has been advanced as a type of phase transition \cite{1}. 
Minimum complexity is a requirement for such a transition, below which only
disconnected subsystems exist, and above which a connected whole emerges.  We
consider here a biological system far less complex, which nevertheless can exhibit 
features of a phase transition in much the same way that the Ising model
illustrates ferromagnetism.  The problem is made concise by focusing on a specific
evolutionary game as a model for the evolution of cooperative behaviour.  The
complexity of the system is not in the basic elements or the dynamics of the
model, but in the variety of spatial patterns possessed by the system.  The notion
of phase transition must be appropriately developed in order to be useful for
biological systems that are not thermal in the physical sense.  The method of
investigation proposed here can be applied to more complex systems.
	
The evolution of cooperative behaviour has been much studied in game theories
with various strategies\cite{2,3}.    Among them the Prisoners' Dilemma is a game
that has received particular attention.  It acquires interesting characteristics when
the game is played in a two-dimensional array\cite{4,5}.  Depending on the payoff
parameter, there can be a rich variety of spatial patterns, ranging from dominance
of one or the other type of players to chaotically changing ones.  Of special interest
is the situation when individuals with one strategy form clusters of all sizes in a
background of those with the other strategy.  Changing the payoff parameter to
either lower or higher values results in the global dominance of one type of
individuals or another.  It is therefore highly interesting to examine the possible
connection that may exist between that state of evolutionary change with phase
transition in physics.
	
Before specifying the problem in detail, it is important to emphasize the
differences that it has from the conventional physical systems under phase
transition.  The latter is generally in thermal equilibrium characterized by a
temperature  $T$.  Below the critical temperature $T_c$ the system exhibits
organized, collective behaviour, while above $T_c$ disordered, thermal motion
dominates.  The behaviour near $T_c$ can be described by critical indices.  None of
these characteristics are valid for the biological system to be studied.  There is
no $T$ and no Boltzmann factor, but there is a payoff parameter $b$.  Instead of
the tension between random and collective modes of motion, the tension is
between the dominance by one and that of the other species, the cooperating and
defecting players.  Instead of a precise $T_c$, there is a range of ``critical'' value
of  $b$.  Instead of the critical indices, it is necessary to find a new exponent that
characterizes the critical behaviour.  Indeed, the concept of phase transition should
be reexamined.
	
To be specific, we consider a particular evolutionary game involving two types of
players labeled  $C$, cooperator, and  $D$, defector.  They occupy the squares of a
two-dimensional lattice with periodic boundary condition, each having eight
neighbours\cite{4,5}.  The initial configuration is that half of the sites are occupied
by  $C$, distributed randomly, and the other half occupied by  $D$.  The
interactions between the occupant of each site with its eight neighbours determine
the configuration of the next time step.  For a $CC$ interaction the payoff is 1 each,
but it is 0 each for a $DD$ interaction.  Between $C$ and $D$, it is 0 for $C$ and $b$
for $D$.  The payoff $b$ is the only tunable parameter in the problem.  Clearly, if
$b$ is high enough, it pays to defect from a $C$ to a $D$.  The actual defection at
each site depends on the total scores of each player when the payoffs of the eight
interactions (with its neighbours) and self-interaction (with itself, assuming
several members of a family occupying the same cell) are added.  When the score
at a site is higher than any of the scores of its neighbours, no change takes place at
that site, reflecting the winning strategy already at hand.  But if the score of any
neighbour is higher and if the highest among them is of the opposite type, then a
switch takes places at the original site to the opposite type at the next time step. 
All such possible switches at all sites are to occur simultaneously and a new
configuration is formed.  The process is then repeated again and again.  
	 
Nowak and May \cite{4,5} have studied the above problem and found many
interesting spatial patterns of $C$ and $D$, depending on the value of $b$.  In fact,
a generalization of the problem to include stochasticity has also been considered
\cite{6}; however, for definiteness only the deterministic one described above is
examined here.  From the point of view of phase transition the most interesting
case is when $b$ is in the range $1.8 < b < 2$, which we shall call the critical range 
$B_c$.  For $b < 1.8$ the lattice is populated mostly by $C$ sites with $D$
scattered in various patterns; for $b > 2.0$ the lattice is almost entirely of $D$ type
with only a few
$C$ sites.  But for $b\in B_c$, clusters of $C$ of various sizes are formed in
a sea of $D$.  The cluster-size distribution has actually been examined in Ref.\
\cite{6}, and is found to decrease slowly without a quantitative description of its
behaviour.  We take the qualitative feature to suggest the possibility of a phase
transition and explore that possibility quantitatively.
	
The first issue to address is that for a biological system considered here there are
no partition function, no thermodynamic functions, no order parameters nor any
other quantity conventionally used in statistical physics to describe phase
transition.  The basic interaction is specified by the rules of the game, and the
dynamical process of type-changing back and forth between $C$ and $D$ at each
site can be simulated on the computer.  The only observables are the locations of
the sites of $C$ at the end of each run, with $D$ being regarded as background. 
This situation is very similar to the production of hadrons in a heavy-ion collision
at very high energy, where the only observables in an experiment are the hadrons
and their locations in momentum space at the end of each collisional event
\cite{7}.  If a quark-hadron phase transition takes place in each of those events, it
is necessary to have a method to infer from the detected hadrons the existence of
such a transition.  Even if the quark system is thermalized before hadronization,
the temperature has no observational relevance, since it is not under experimental
control.  Thus the problem of determining the signature of phase transition is very
similar in the biological and quark systems. 
	
For quark-hadron phase transition the Ginzburg-Landau theory has been used to
examine the scaling properties of the normalized factorial moments \cite{8}, which
are defined by \cite{9}
	\begin{eqnarray} 
F _q (\delta)= \left<{\left< n(n-1)\cdots(n-q+1)\right>_{\delta} \over \left<n
\right>^q_\delta}\right>_e
\quad ,
\label{1} 
     \end{eqnarray} 
where $n$ is the particle multiplicity in a bin of size  $\delta$,  $\left<
\cdots \right>_\delta$ is the average over all bins of the same size in an event,
and $\left< \cdots \right>_e$ is the average over all events.  The virtue
possessed by $F_q$  is that it filters out statistical fluctuations so that only
dynamical fluctuations contribute to nontrivial values of $F_q$.  At the critical
point when hadrons are produced, one expects the spatial pattern of the produced
particles in phase space to exhibit self-similar behaviour, which ideally (as in the
Ising model
\cite{10}) would imply a power-law dependence of $F_q(\delta)$ on $\delta$, i.e.
\cite{9,11}
   \begin{eqnarray} 
F _q (\delta) \propto \delta^{-\varphi _q}  \label{2} 
     \end{eqnarray}
for a range of $\delta$ .  Strict power-law behaviour for $F_q$ is, however, not
found in the Ginzburg-Landau theory \cite{8}.  Instead, $F$-scaling in the form
	\begin{eqnarray} 
F_q \propto F^{\beta _q}_2 \quad , \label{3} 
     \end{eqnarray}  
as $\delta$ is varied, is found to be well satisfied.  Furthermore, the slopes $\beta
_q$ in the log-log plot are found to be extremely well fitted by the formula
	 \begin{eqnarray} 
\beta _q = (q - 1)^{\nu}  \label{4} 
     \end{eqnarray}	
where  $\nu$, called the scaling exponent, is found to be 1.304.  This value for
$\nu$ is independent of the details of the Ginzburg-Landau free energy or the
critical temperature.  It has therefore been suggested as a signature for phase
transition, when the only observable is the particle multiplicity \cite{8}.
	
So far no experiments in heavy-ion collisions have produced data that yield 
$\nu = 1.304$.  All such experiments on hadron production have resulted in $\nu$ 
greater than 1.55.  It means that according to the criterion of Ref.\ \cite{8} no
quark-gluon plasma has been formed before hadronization in those collisions.  The
verification of $\nu = 1.304$ has, however, been found in quantum optics where
an experiment on photon-number fluctuations near the threshold of lasing (which
is known to be describable by a second-order phase transition) measured the
$F_q$ moments that satisfy (\ref{3}) and (\ref{4}) with high degree of accuracy
\cite{12}.
	
The above procedure of analyzing spatial fluctuations is eminently suitable for the
evolutionary game problem, where a $C$ site may be regarded as a particle in a
two-dimensional phase space.  We have made computer simulations of the
evolution on a $120 \times 120$ lattice, starting with 50\% $C$ sites distributed
randomly on the lattice. After 50 time steps we  count the number of $C$
sites in bins of size $\delta ^2$, where $\delta$ is varied from 5 to 20.  After
simulating
$10^3$ times, the $F_q$  moments are calculated. The result is shown in Fig.\ 1 as a
function of the number of bins in the lattice, $M = (120/\delta)^2$.  The log-log
plot reveals almost linear rise, although not in exactly straight lines.  Thus the
scaling behaviour in (\ref{2}) is only approximately satisfied, but sufficiently so to
suggest that the clusters of $C$ sites exhibit features reminiscent of critical
behaviour.  This behaviour is found only for $b \in B_c$, which we have referred to
as the critical region, principally because scaling behaviour is the characteristic of
criticality.
	
To see how close the behaviour matches the Ginzburg-Landau type of phase
transition, we plot log$F_q$ versus log$F_2$ in Fig.\ 2 for $q = 3, \cdots, 6$. 
Evidently, for lower $q$ values the points align very nearly on straight lines, but
at higher $q$ values saturation develops.  The local slopes can be measured and
plotted as a range of $\beta _q$  for each $q$, shown in Fig.\ 3.  The log$\beta _q$
versus log$(q-1)$ plot shows a linear dependence in good agreement with
(\ref{4}).  The value of the scaling exponent is determined to be
   \begin{eqnarray} 
\nu = 1.23 \pm 0.15 \quad . \label{5} 
     \end{eqnarray}	 
This is to be compared with the value of 1.304 derived from the Ginzburg-Landau
theory \cite{8}.
	
There is, of course, no reason why the evolutionary game that we have considered
should fit into the Ginzburg-Landau description of phase transition.  Nevertheless,
the fact that (\ref{5}) agrees with 1.3 within its errors gives quantitative support
to the suggestion that when the payoff parameter $b$ is increased to the critical
range $B_c$, the change from cooperators $C$ to defectors $D$ corresponds to
a phase transition.  It is worth repeating that no experiments of any kind, except
for lasers at threshold \cite{12}, have produced a value of $\nu$ close to 1.3.  It
is also pertinent to remark that for Ising model in two dimensions $\nu$ has
recently been found to be between 1.0 and 1.5, depending on how close the spin
system is to $T_c$ \cite{10}.  Since the Ising model is a precise example of phase
transition, and resembles more closely the evolutionary game than
Ginzburg-Landau by virtue of the commonality in 2D lattice and the
near-neighbour nature of the interactions, the value of $\nu$  in (\ref{5}) sits
comfortably in the range to be indicative of phase transition.
	
We have used the factorial moments $F_q$ and the scaling exponent $\nu$ to
determine the scaling behaviour of the distribution of the cooperators $C$ in the
two-dimensional space.  It is then inferred that for the payoff  $b$  in the critical
range the system undergoes a phase transition.  At this stage we know of no
alternative method by which the issue of phase transition can be examined for
such a biological system.  This type of investigation should be extended to include
stochastic possibilities of the evolutionary games and multistate generalization of
cooperation.

There are several implications to be derived from the result of this study. 
Firstly, the concept of phase transition, which has largely been overlooked in
biology, is now given further concrete support for relevance.  One of the
characteristics of the conventional critical phenomena is the scaling behaviour in
cluster sizes.  Its implication for evolutionary biology is therefore that one should
be aware of cluster formation in the gathering of field data.  For theoretical biology
this result suggests the possibility of universality classes of evolution dynamics,
since the hallmark of critical phenomena is their universality.  Of more immediate
significance is the realization that the range of $\nu$ (between 1.2 and 1.3)
characterizing phase transition is so universal in its validity that the systems
embraced by it as observable signature range from the subnuclear (quark-gluon
plasma), to the atomic (lasers), to condensed-matter (ferromagnetism), and now to
the biological  (evolutionary games).

\begin{center}
\subsection*{Acknowledgment}
\end{center}

One of us (RCH) is grateful to R.\ M.\ May for communications that stimulated this
research.  Discussion and consultation with P.\ Bak and M.\ A.\ Nowak have also
been helpful.  This work was supported in part by the U.\ S.\ Department of
Energy under Grant No.\ DE-FG06-91ER40637.

\begin{center}
\subsection*{Figure Captions}
\end{center}

\begin{description}
\item[Fig. 1.]  On a lattice of size $L^2$ the numbers of cooperators $C$ are
counted in bins of size $\delta^2$  and then used in the calculation of the factorial
moments  $F_q$.  This figure shows the dependence of $F_q$ on the number of
bins $M = (L/\delta)^2$  for various fixed values of $q$.

\item[Fig. 2.]	The $F_q$ moments exhibit approximate  $F$-scaling in accordance
to the power-law behaviour shown in Eq.\ (\ref{3}).

\item[Fig. 3.]	The scaling exponent $\nu$ as defined by $\beta_q = (q-1)^{\nu}$ is
shown to be $1.23 \pm 0.15$.
\end{description}

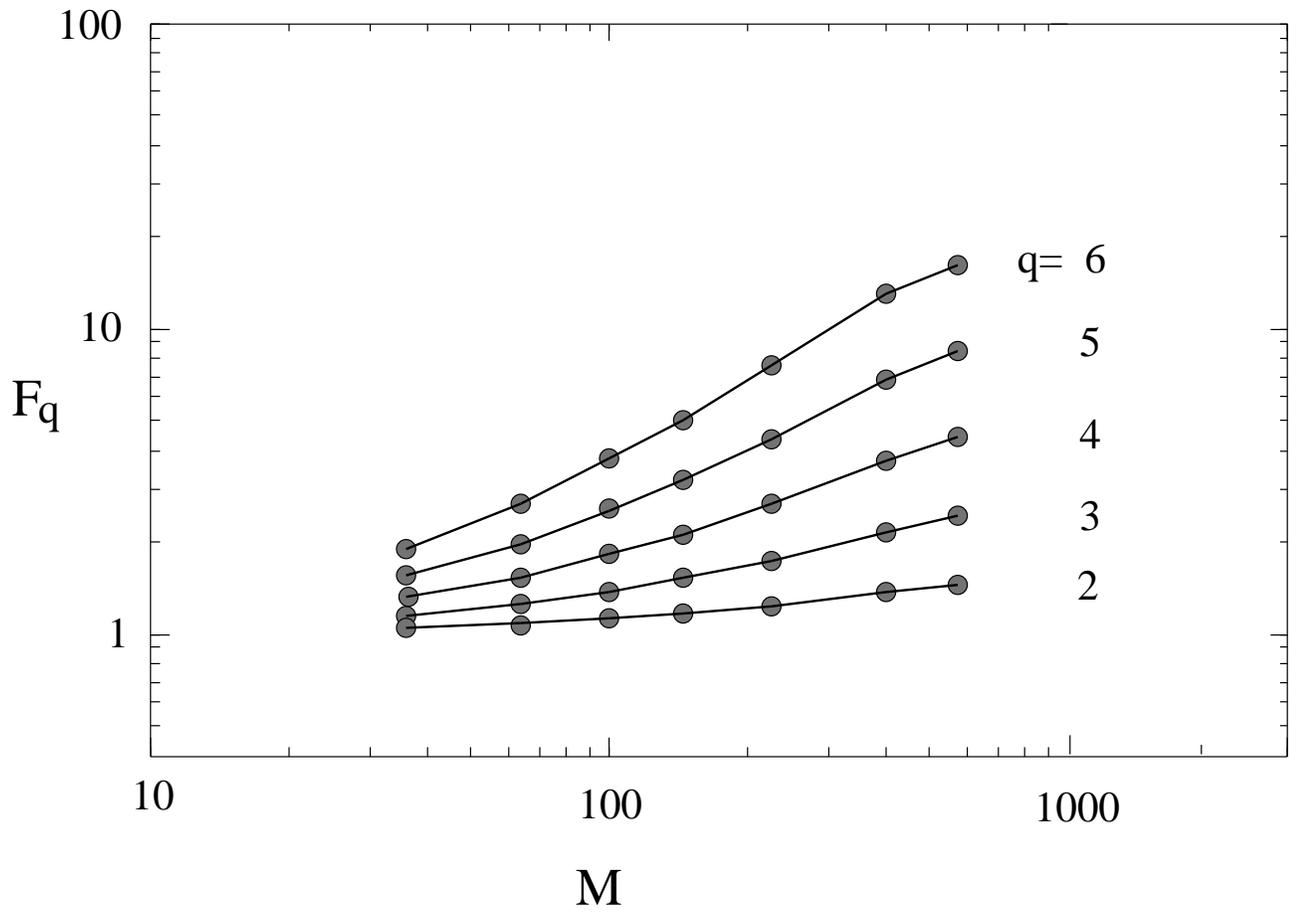
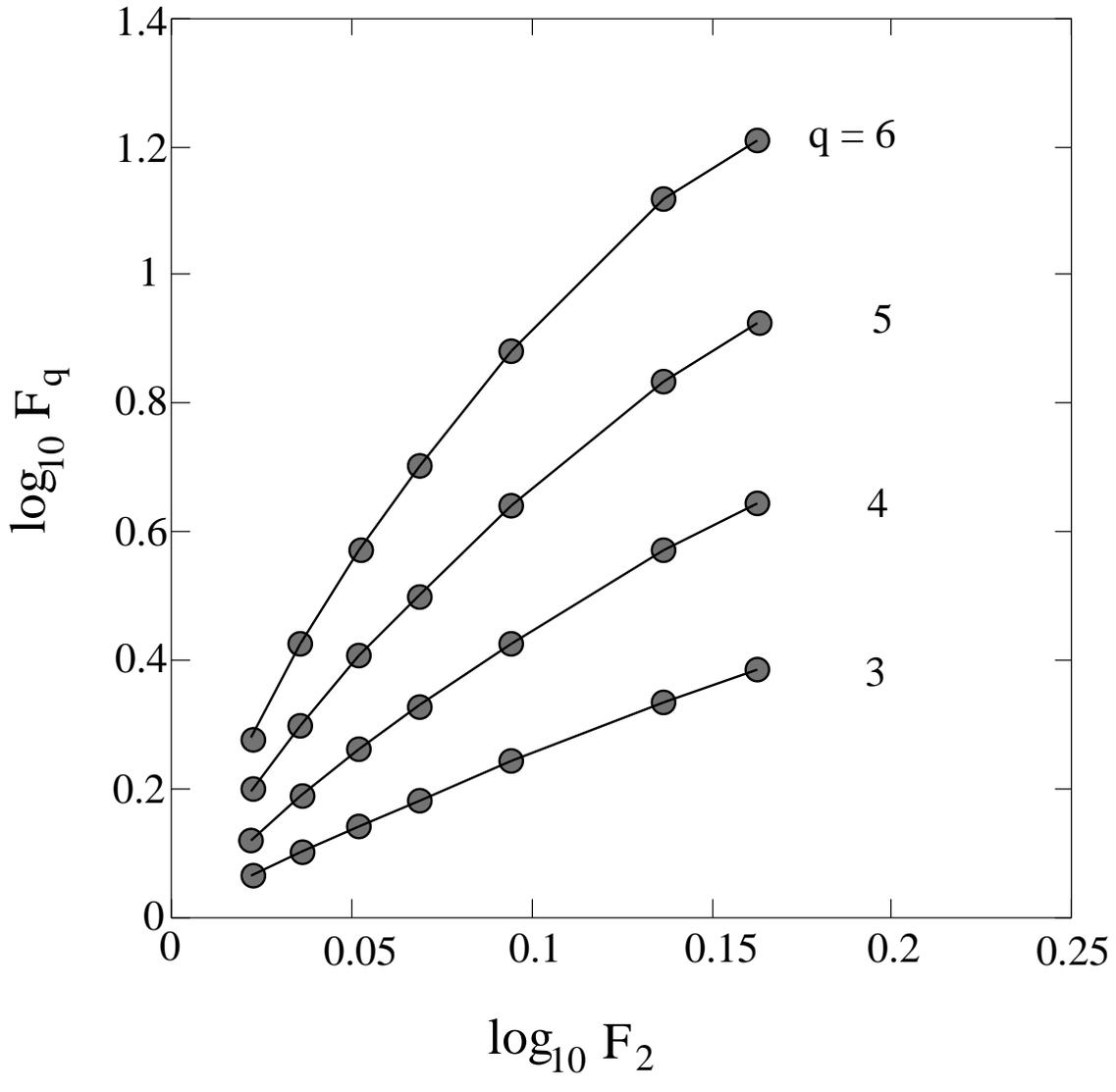
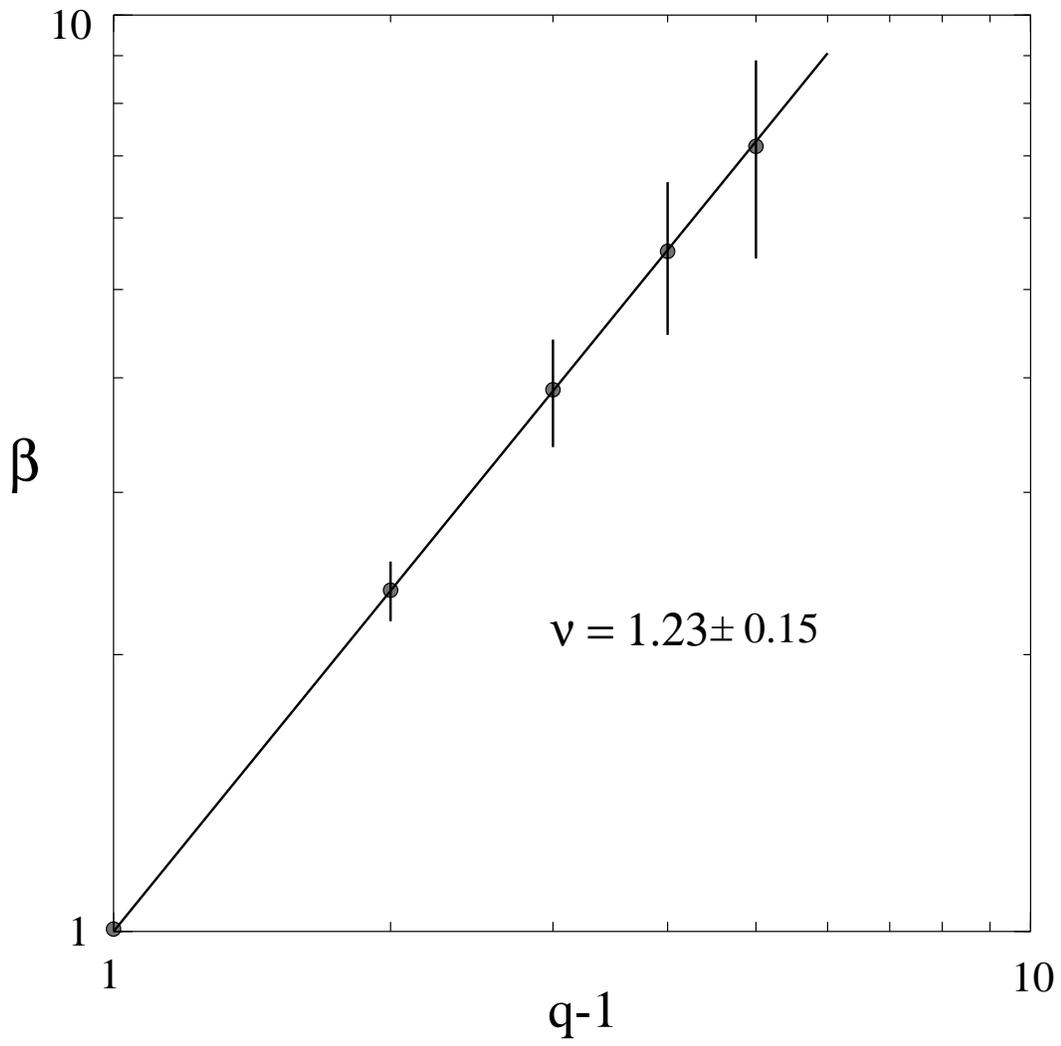
\end{document}